\documentclass{article}
\usepackage{spconf,amsmath,graphicx}
\usepackage{hyperref}
\usepackage{subcaption}
\usepackage{multirow}
\usepackage{pgfplots}
\usepackage{tikz}
\usepackage{paralist}
\usepackage{xcolor}
\usepackage{fixltx2e}
\usetikzlibrary{
    patterns,
}
\usepgfplotslibrary{
    groupplots,
}

\title{Audio ALBERT: A Lite BERT for Self-supervised Learning of Audio Representation}
\name{Po-Han Chi\begin{math}^1\end{math}, Pei-Hung Chung\begin{math}^1\end{math}, Tsung-Han Wu\begin{math}^1\end{math},Chun-Cheng Hsieh\begin{math}^1\end{math}, Yen-Hao Chen\begin{math}^1\end{math},  \\ \emph{Shang-Wen Li}\begin{math}^{2}\end{math},\sthanks{work done before joining Amazon}
\emph{Hung-yi Lee}\begin{math}^{1}\end{math} 
}

\ninept

\address{
  \begin{math}^{1}\end{math}College of Electrical Engineering and Computer Science, National Taiwan University \\
  \begin{math}^{2}\end{math}Amazon AI
  }
\pgfplotsset{compat=1.16}
\begin{document}

\maketitle
\begin{abstract}

Self-supervised speech models are powerful speech representation extractors for downstream applications. Recently, larger models have been utilized
in acoustic model training to achieve better performance. We
propose Audio ALBERT, a lite version of the self-supervised
speech representation model. We apply the light-weight representation extractor to two downstream tasks, speaker classification and phoneme classification. We show that Audio ALBERT achieves performance comparable with massive pre-trained networks in the downstream tasks while having 91\%
fewer parameters. Moreover, we design probing models to
measure how much the latent representations can encode the
speaker’s and phoneme’s information. We find that the representations encoded in internal layers of Audio ALBERT contain more information for both phoneme and speaker than the
last layer, which is generally used for downstream tasks. Our findings provide a new avenue for using self-supervised networks to achieve better performance and efficiency. 
\end{abstract}

\section{Introduction}
\label{sec:intro}

Recently, pre-trained models~\cite{devlin2019bert, peters2018deep, radford2018improving, radford2019language}, especially BERT,  dominate Natural Language Processing (NLP) world. 
The models learn powerful and universal representation by utilizing self-supervised learning at the pre-training stage to encode the contextual information. 
The representation is beneficial to performance, especially when the data of the downstream task is limited. As of late, BERT-like models are also applied to the speech processing domain.
The pre-trained model learns the robust speech representations for speech processing tasks, such as  Automatic Speech Recognition (ASR) and speaker recognition, with the self-supervised learning.
approaches~\cite{liu2020mockingjay, jiang2019improving, ling2020deep, Baskar2019, Schneider2019}.

However, since the size of these BERT-like pre-trained models is usually prohibitively large, these models require a significant amount of memory for computation, even at the fine-tuning stage. 
The requirement hinders the application of pre-trained models from different downstream tasks.  

ALBERT~\cite{lan2019albert} addresses the challenge of efficiency. ALBERT is a lite version of BERT for text by sharing one layer parameters across all layers and factorizing the embedding matrix to reduce most parameters. Although the number of parameters is reduced, the representations learned in ALBERT are still robust and task agnostic, such that ALBERT can achieve similar performance in the same downstream tasks comparing to BERT. 

In this paper, we first examine the knowledge encoding in each layer of Mockingjay, a pre-trained model utilizing BERT architecture to encode speech information. We found the learned parameters are redundant across layers. Thus, we bring the idea of sharing parameters from ALBERT to the speech processing domain and propose a novel self-supervised model, Audio ALBERT (AALBERT), for parameter-efficient representation learning.

We show that AALBERT yields comparable performance to other pre-trained models in downstream tasks, but with much smaller networks. To understand how to use the pre-trained networks properly in downstream tasks, we also analyze representations extracted from different layers of AALBERT. We use a simple classifier to probe each layer, and we find that the representations of the intermediate layers contain more phonetic and speaker information than that of the last layer. The finding indicates that the representations from the last layer fit the pre-training task too much, and the intermediate layers may be more suitable for adapting to downstream tasks. To our best knowledge, this is the first study to bring the idea of model compression in ALBERT to speech processing, to show the benefits in the efficiency of the novel architecture, AALBERT, for speech-related tasks, and to analyze learned latent representations for better usage of pre-trained networks in downstream tasks. The code will be available soon (\url{https://github.com/pohanchi/AALBERT})

\section{Related work}

\subsection{Self-supervised learning representation}
\label{sec:relate_work}

In recent years, works related to self-supervised learning spring up in Computer Vision (CV), NLP, speech processing, etc.
In CV, some works~\cite{chen2020simple, he2020momentum} incorporate contrastive objective and self-supervised learning for learning visual representation. Self-supervised learning is also utilized to learn language representations for NLP tasks. ELMo~\cite{peters2018deep} is the first work introducing the concept of contextualized embeddings and the weighted sum application. BERT~\cite{devlin2019bert} further presents the concept of Masked Language Model (MLM). Deep transformer encoder architecture is trained with MLM to reconstruct the masked input sequences in the pre-training stage. The resulting networks show substantial performance gain in downstream NLP tasks. XLNet~\cite{yang2019xlnet}, introduces the Permutation Language Model and outperforms both autoregressive models and MLM. However, Roberta~\cite{liu2019roberta}, achieves performance comparable with XLNet by training with more data, larger batch size, and the better hyperparameter settings. For parameter efficiency, ALBERT~\cite{lan2019albert} is proposed to reduce the model size without losing performances in NLP tasks compared to BERT.
Self-supervised learning is also gaining attention in the speech field. 
Contrastive Predictive Coding (CPC)~\cite{oord2018representation} incorporates contrastive objective in self-supervised learning to learn powerful representations for speech processing tasks.  
Autoregressive Predictive Coding (APC)~\cite{Chung2019} utilizes the idea of an autoregressive model from ELMo to learn stronger speech representations. 
Inspired by MLM, Mockingjay~\cite{liu2020mockingjay} masks frame from input acoustic feature and pre-trains the networks to reconstruct the corresponding linear spectrogram or mel spectrogram in the pre-training stage. 
Similarly, Masked Predictive Coding (MPC)~\cite{jiang2019improving} uses the idea of MLM to pre-train a model for speech recognition. 
Speech-XLNet~\cite{song2019speech} is the audio version of XLNet. vq-wav2vec~\cite{baevski2019vq} incorporates vector quantization and BERT to improve the performance on downstream tasks. 
Finally, DeCoAR~\cite{ling2020deep}, a model built with a deep LSTM module, adopts pre-training tasks similar to Mockingjay and MPC and yields significant performance gain in speech recognition. 
All of these pre-trained networks are large in model size and focus on improving performance with more parameters or pre-training data. 
To make the models more compact for training and deployment, we build a lite version of a pre-trained network that yields comparable performance with fewer parameters and memory footprint.

\subsection{Weight sharing}
    The previous work~\cite{press2017using} ties the input and output embeddings to reduce parameters without harming the performance.
    Tong Xiao et al. proposes a method~\cite{DBLP:journals/corr/abs-1906-11024}, which reuses the attention weights of previous layers in the adjacent layers on the transformer model for faster inference and keeps performance in neural machine translation.    
    Some works build compact transformer models~\cite{dabre2019recurrent, lan2019albert}, which apply weight sharing mechanisms across layers to reduce parameters and achieve comparable performance in their tasks.
    Dehghani et al. proposed Universal Transformer~\cite{dehghani2019universal}, which utilizes the benefit of the transformer and recurrent neural network, and it also incorporates weight sharing across layers to reduce a great number of parameters but keep performance in the different tasks. 
    In general, weight sharing mechanism across layers can be viewed as an RNN applied in the direction of the layer-axis. 
    To sum up, weight sharing can not only bring faster inference and training speed but keep similar performance in previous works. 
    It is a kind of network-compression mechanism to reduce parameters heavily in transformer models.

 \subsection{Probing task}
 Probing is a technique to measure whether the encoder embeds specific information in representation~\cite{jawahar2019does, li2020does, Belinkov2019}. 
The probing is done by extracting representation to be examined, building a simple classifier based on the representation for a downstream probing task, and measuring the classifier's performance.
Synthesizing audio from the ASR hidden state is also proposed~\cite{li2020does}, as another way of probing.

\section{Method}
    \subsection{Mockingjay}
    \label{sec:mockingjay}
        Mockingjay~\cite{liu2020mockingjay} is a pre-trained model that utilizes the architecture of BERT model. In pre-training, Mockingjay takes a masked spectrogram as the input to reconstruct the original one. There are three common model architectures for Mockingjay by adopting 3, 6, and 12 layers of transformer encoders (denoted as Mockingjay-3L, Mockingjay-6L, and Mockingjay-12L respectively). The previous study
       ~\cite{liu2020mockingjay} showed that the representation of Mockingjay possesses both rich phonetic and speaker information.  
       
    \begin{figure}[t]
        \centering
        \begin{subfigure}{.45\columnwidth}
            \includegraphics[width=\textwidth]{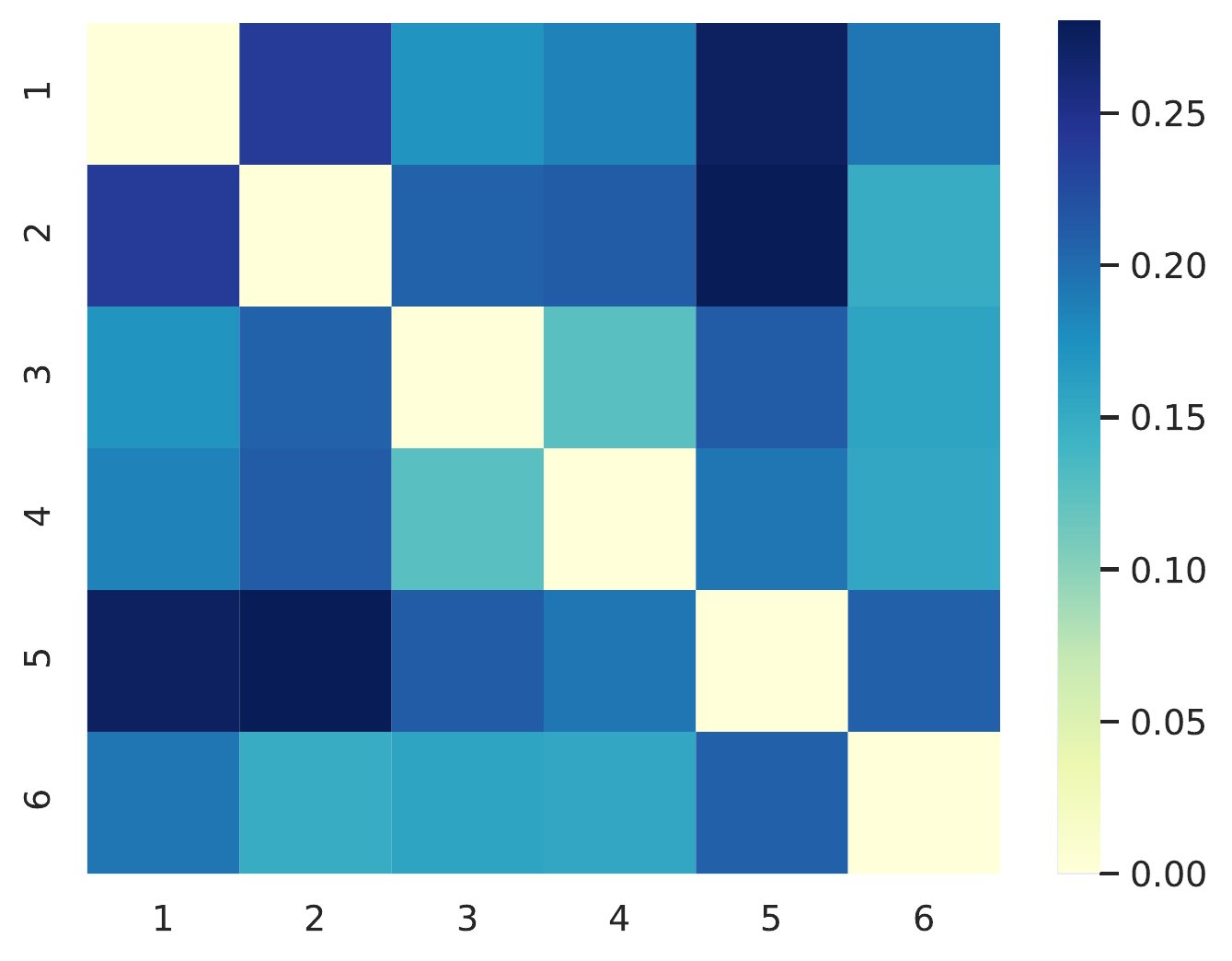}
            \caption{Average}
            \label{fig:js_attention_mocking}
        \end{subfigure}%
        \begin{subfigure}{.45\columnwidth}
            \begin{minipage}[t]{.5\columnwidth}
                \includegraphics[width=\textwidth]{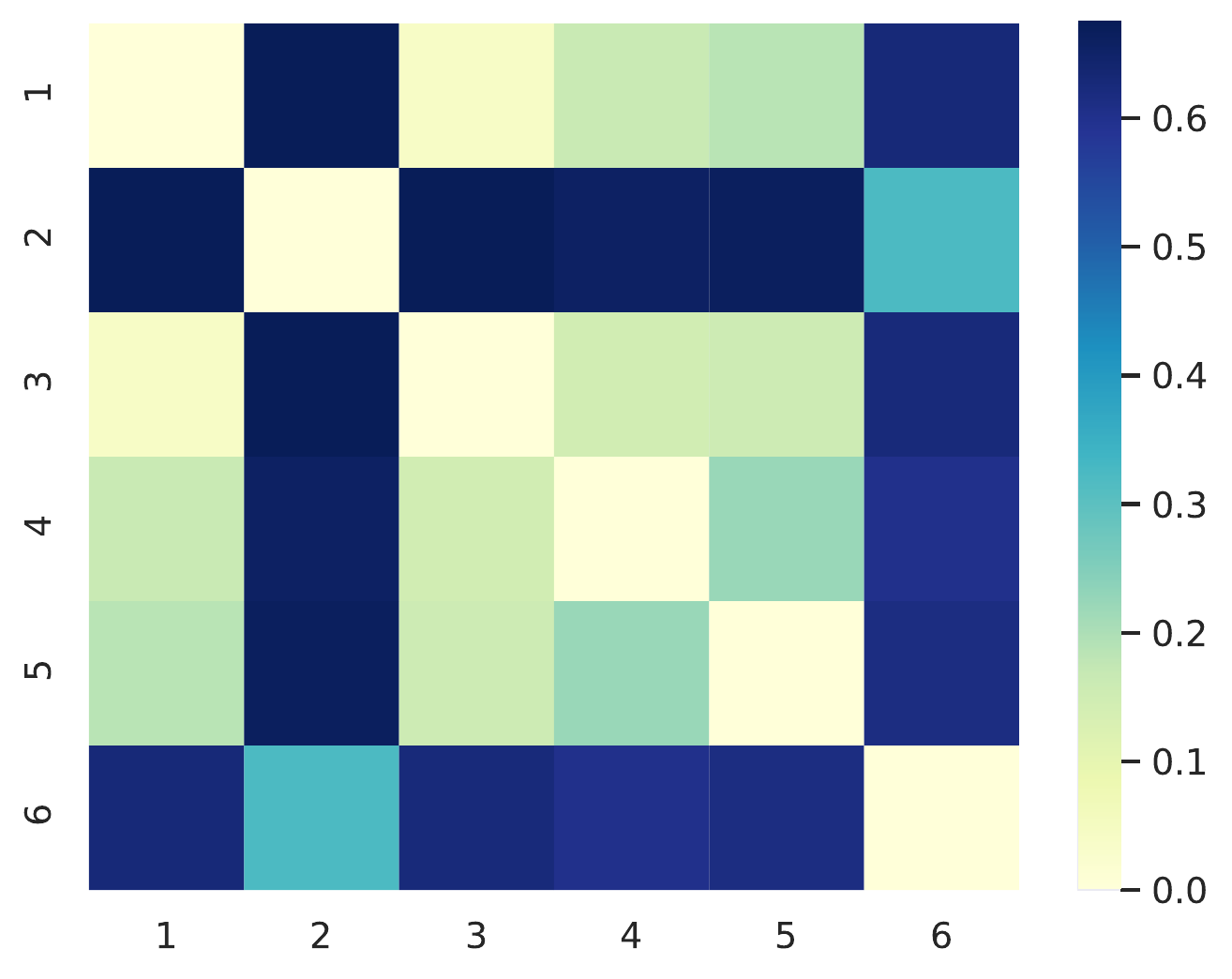}
                \caption{}
                \label{fig:js_attention_mockingjay_1}
            \end{minipage}%
            \begin{minipage}[t]{.5\columnwidth}
                \includegraphics[width=\textwidth]{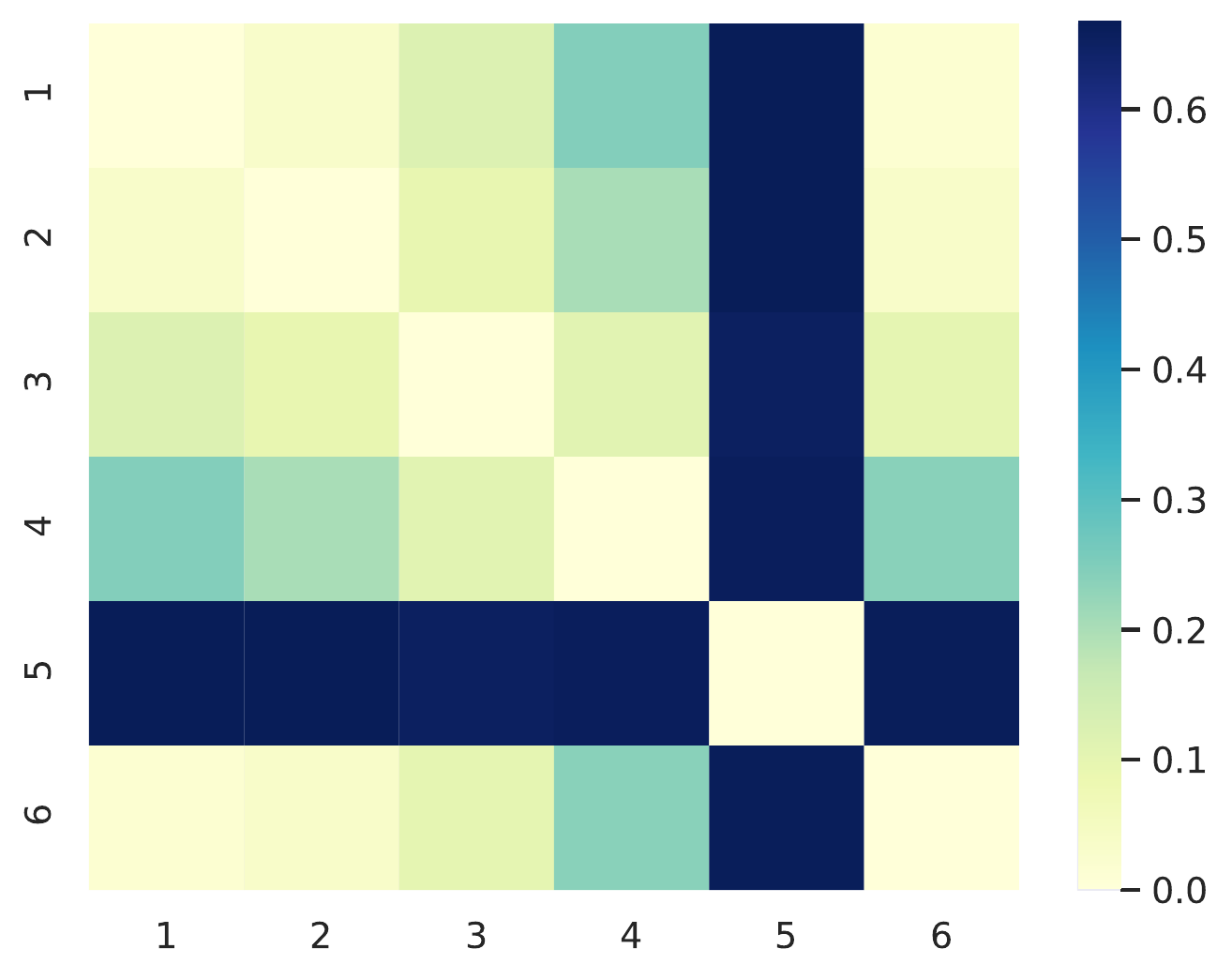}
                \caption{}
                \label{fig:js_attention_mocking_2}
            \end{minipage}
            \begin{minipage}[t]{.5\columnwidth}
                \includegraphics[width=\textwidth]{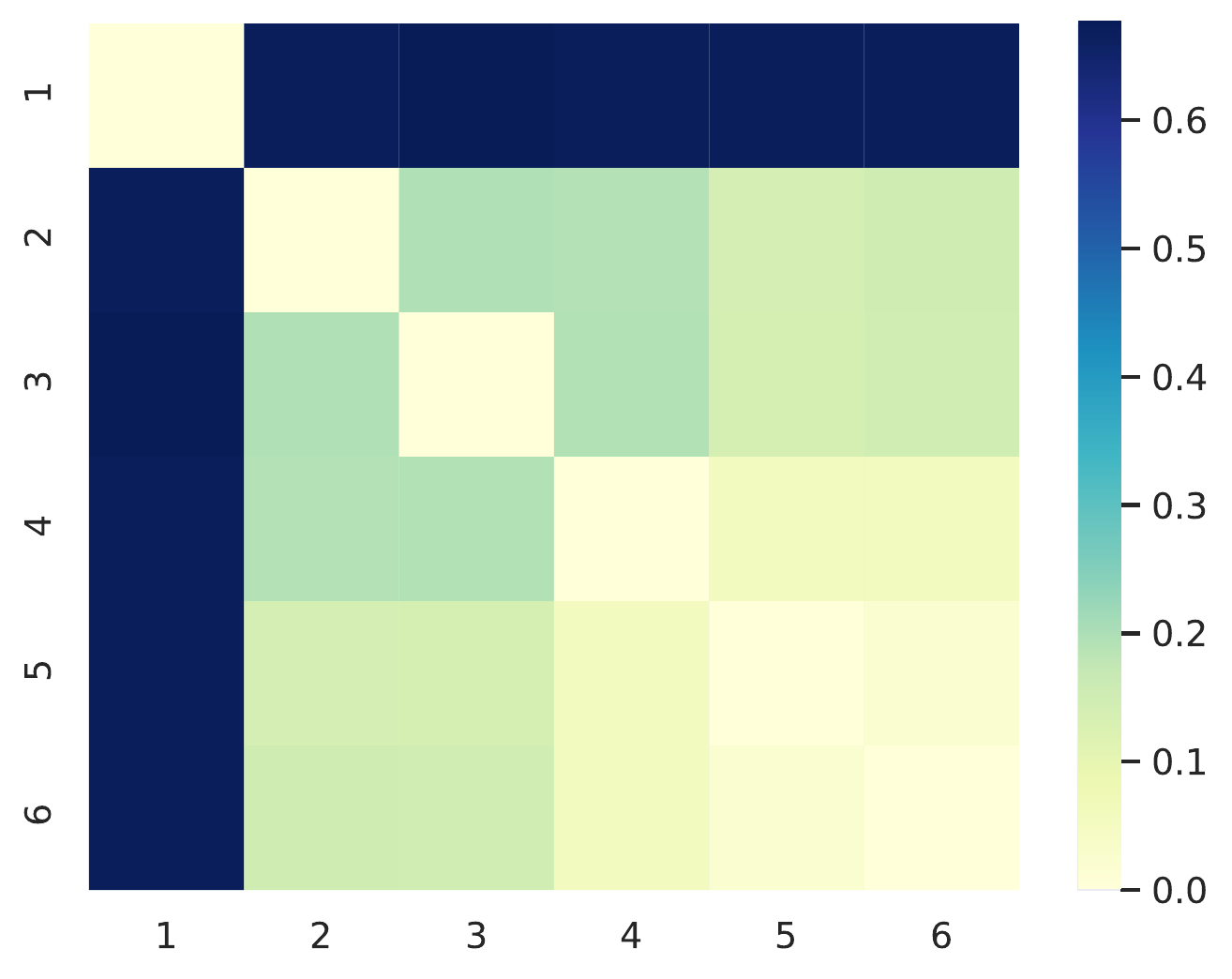}
                \caption{}
                \label{fig:js_attention_mocking_3}
            \end{minipage}%
            \begin{minipage}[t]{.5\columnwidth}
                \includegraphics[width=\textwidth]{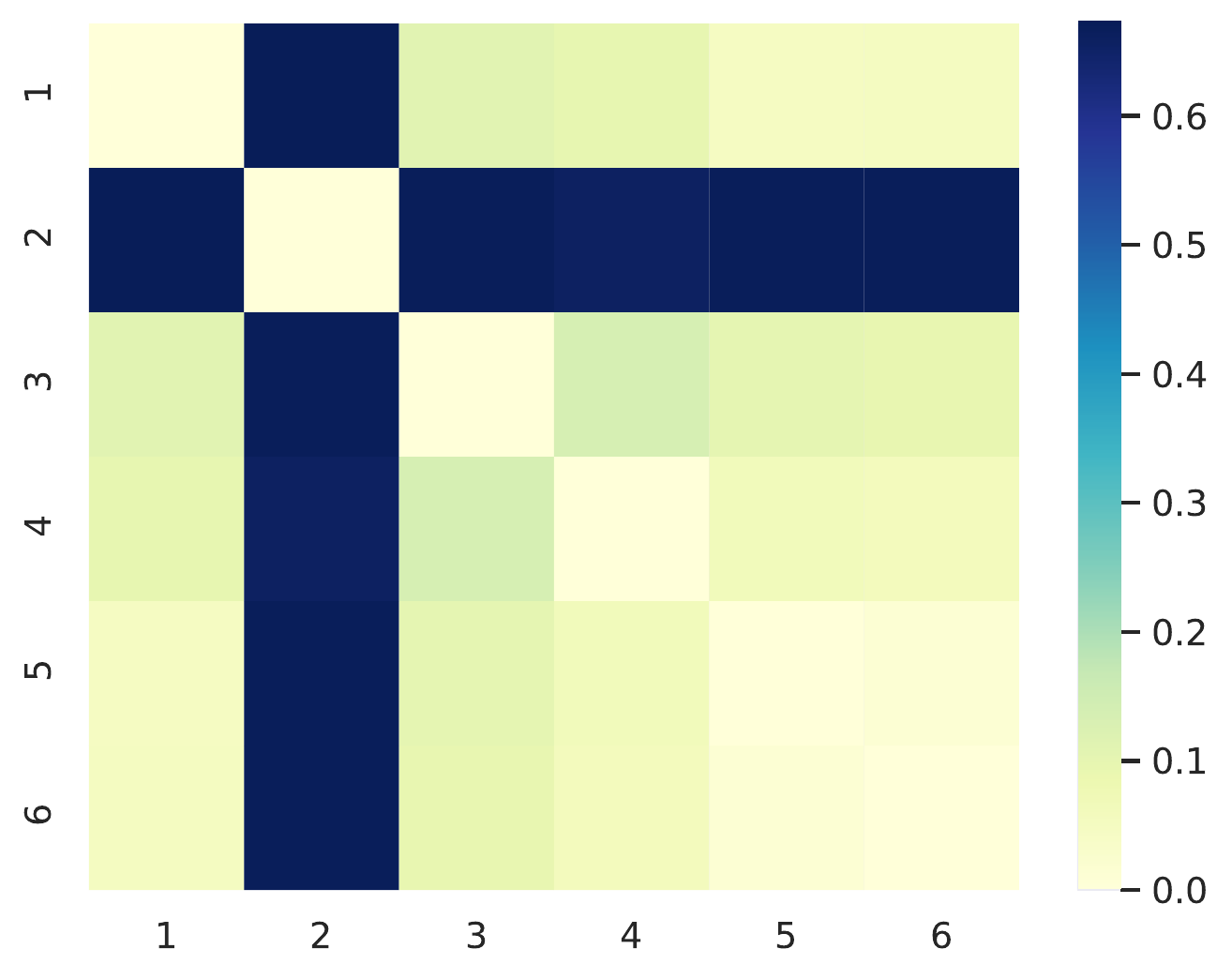}
                \caption{}
                \label{fig:js_attention_mockingjay_4}
            \end{minipage}
        \end{subfigure}
        \caption{The JS divergence of attention distribution between different layers in Mockingjay-6L. (a) is the average case, while (b)(c)(d)(e) represent different attention heads.}
    \end{figure}
    
        We further investigate the parameter usage in Mockingjay (due to space limitations, we only show results for the most commonly used architecture, Mockingjay-6L. Our findings here also apply to the other two variations).
        Inspired by Fast Transformer~\cite{DBLP:journals/corr/abs-1906-11024} Experiments, we use the Jensen-Shannon (JS) divergence to evaluate the difference between the attention distribution of each transformer encoder layer. For the multi-head attention, we calculate the JS divergence within each head, then average them to obtain the JS divergence between every layer. 
        Figure~\ref{fig:js_attention_mocking} shows the JS divergence of attention distribution between layers in Mockingjay-6L.
        Here we can see that JS divergence between layers in Mockingjay-6L is significant.
        Furthermore, 
        we randomly pick one attention head every layer and do the same experiment. We show them in
        Fig~\ref{fig:js_attention_mockingjay_1}, Fig~\ref{fig:js_attention_mocking_2}, Fig~\ref{fig:js_attention_mocking_3}, and Fig~\ref{fig:js_attention_mockingjay_4}. Although some attention heads are very different from those in other layers (dark blue cells), most layers are similar (light blue cells). The result shows that for a specific attention head in Mockingjay-6L, there is usually some similar attention distribution over different layers. 

        Although the parameters of each layer are different, they still generate similar attention distribution.
        This phenomenon indicates the existence of redundancy in parameter usage and the possibility to compress models via weight-sharing across layers without sacrificing model expressiveness.

\begin{figure}[t]
    \centering
    \includegraphics[width=.85\linewidth]{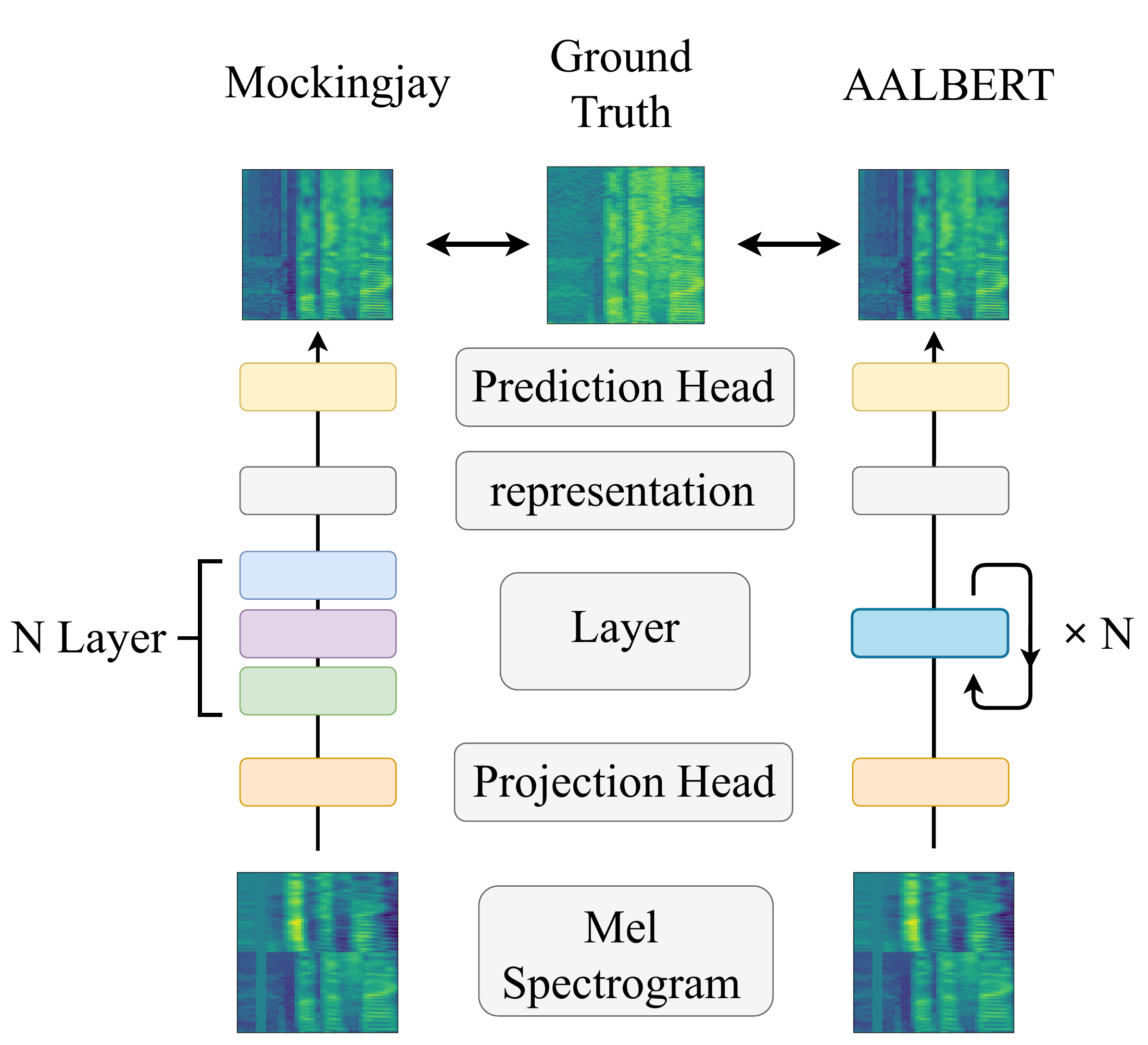}
    \caption{Difference between Mockingjay and AALBERT}
    \label{fig:diff}
\end{figure}

\begin{table}[t]
    \caption{pre-trained Models}
    \label{tab:pre-trained model }
    \centering
    \resizebox{0.95\columnwidth}{!}{
    \begin{tabular}{l|ccc}
      \hline
      \textbf{Model}      & \textbf{Layer} & \textbf{Params} & \textbf{Param Sharing}               \\
      \hline
      AALBERT-12L              & 12      &  7.4M       &   True              \\
      AALBERT-6L           & 6       &  7.4M       &   True              \\
      AALBERT-3L           & 3       &  7.4M       &   True              \\
      \hline
      Mockingjay-12L           & 12      &  84.3M      &  False               \\
      Mockingjay-6L        & 6       &  44.4M      &  False                \\
      Mockingjay-3L        & 3       &  21.6M      &  False                \\
      \hline
    \end{tabular}
    }
\end{table}

    \subsection{AALBERT}
        We propose AALBERT, or audio ALBERT, for a more compact pre-trained network. Similar to Mockingjay, AALBERT also takes mel-spectrogram as the input acoustic features. We mask the input features with zero and pre-train the network to reconstruct the corresponding log-linear spectrogram after applying Cepstral Mean and Variance (cmvn) from the masked input. We apply the masking to features of each utterance by first downsampling one out of every three frames and then randomly selecting 15\% of the resulting frames for adding noises. The noises are introduced as the following. We zero out 80\% of the selected frames, replace the frames with other frames randomly sampled from the same utterance with 10\% probability, and keep the original frames for the remaining cases.
        
        As compared to Mockingjay, we introduce weight tying for reducing parameters. As visualized in Fig~\ref{fig:diff}, both Mockingjay and AALBERT are built with the architecture of the Transformer encoder, where each layer of the encoder consists of components including self-attention, feed-forward, and layer normalization. However, in AALBERT, parameters of each component are shared over all the layers. To illustrate the gain in efficiency, in Table 1, we compare the numbers of parameters for all pre-trained models studied in this paper. As we can see, AALBERT requires much fewer parameters than Mockingjay for the same depth of Transformer. All our experiments are modified from Mockingjay Large model, which is 12 layers and utilizes linear spectrogram reconstruction as a pre-training task. We do not use the Mockingjay base model because we want to compress the size of the original model, and the base model is only 3 number of layers, which limit the space of compression.

\begin{figure}[h]
    \centering
    \begin{subfigure}[t]{.99\columnwidth}
    \includegraphics[width=\linewidth]{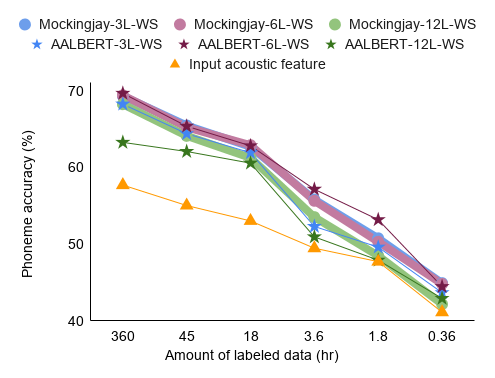}
    \caption{Feature-extraction case}
    \label{fig:weightedsum}    
    \end{subfigure}
    \begin{subfigure}[t]{.99\columnwidth}
    \includegraphics[width=\linewidth]{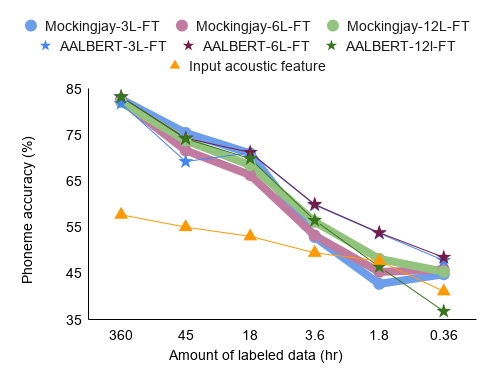}
    \caption{Fine-tuning case}
    \label{fig:finetune}
    \end{subfigure}
    \caption{Phoneme classification accuracy vs amount of labeled data. 3L, 6L, 12L: the number of layers, FT: fine-tune, WS: weighted sum, Input acoustic feature: input acoustic feature as baseline.}
\end{figure}

\begin{figure*}[t]
    \centering
    \begin{tikzpicture}
        \begin{axis}[
            xlabel={Number of layers},
            ybar,
            ymin=55,
            ymax=94,
            y label style={at={(axis description cs: 0.475,1.64)},anchor=center,rotate=-90},
            ylabel={Number of million parameters of AALBERT (A) and Mockingjay (M) },
            extra y ticks={75.5},
            extra y tick labels = { Accuracy (\%) },
            extra y tick style = {
                            tick label style={
                                            rotate=90,
                                            yshift=7mm,},
                            },
            bar width=0.55cm,
            enlarge x limits=0.25,
            height=3.7cm,
            legend cell align={left},
            legend style={
                            at={(0.5,-0.7),
                                },
                            draw=none,
             /tikz/every even column/.append style={column sep=5pt},
              anchor=north,legend columns=2},
              transpose legend,
            symbolic x coords={3, 6, 12},
            extra x ticks={3, 6, 12},
            extra x tick labels={{7.4 \textsubscript{(A)} 22.6 \textsubscript{(M)}}, {7.4 \textsubscript{(A)}, 44.4 \textsubscript{(M)}}, {7.4 \textsubscript{(A)}, 84.3 \textsubscript{(M)}}},
            extra x tick style={
                    ticklabel pos=top,
                },
            xtick=data,
            x=4.5cm,
            nodes near coords,
            nodes near coords align={vertical},
                    nodes near coords style={
                            font=\footnotesize,
                            /pgf/number format/.cd,
                            fixed,
                            fixed zerofill,
                            precision=1,
                        },
            ]
            \addplot [pattern=crosshatch dots, pattern color=blue,]
            	coordinates {(3, 65.67) (6, 68.87) (12, 61.86)};
            \addplot [pattern=crosshatch dots, pattern color=red,]
            	coordinates {(3, 67.13) (6, 64.05) (12, 58.16)};
            \addplot [fill=cyan!50!white]
            	coordinates {(3,68.5) (6, 69.97) (12, 67.26)};
            \addplot [fill=orange!80!white]
            	coordinates {(3, 69.31) (6, 69.39) (12, 68.23)};
            \addplot [fill=teal!60!white,]
            	coordinates {(3, 81.79) (6, 83.21) (12, 83.21)};
            \addplot [fill=red!75!white,]
            	coordinates {(3, 82.72) (6, 82.37) (12, 82.08)};
        \legend{last-layer AALBERT , last-layer Mockingjay, weighted-sum AALBERT, weighted-sum Mockingjay, finetuned AALBERT, finetuned Mockingjay}
        \end{axis}
    \end{tikzpicture}
    \caption{Phoneme classification accuracy}
    \label{fig:phoneme_classification}
\end{figure*}
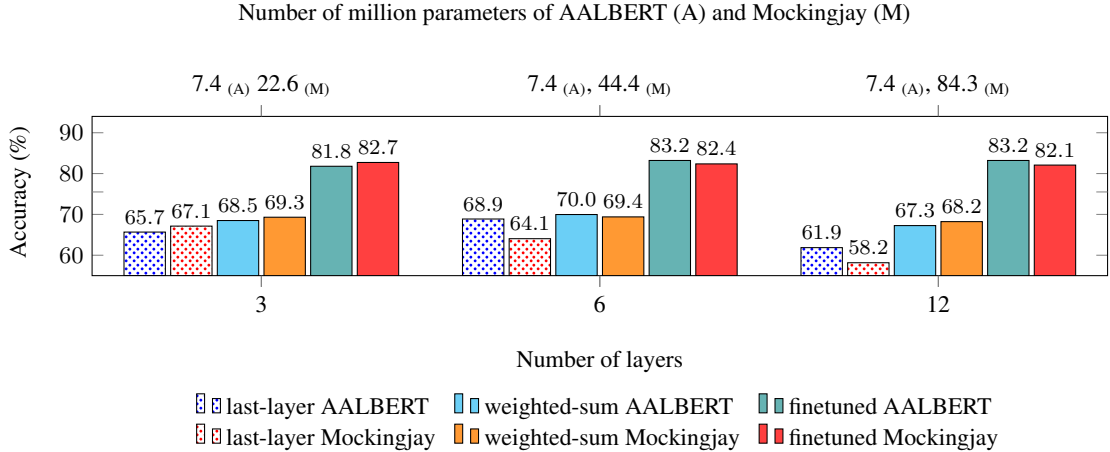

    \subsection{Application to downstream tasks}
        We investigate two popular ways of applying pre-trained networks to downstream tasks: feature extraction and fine-tuning.
    
        \subsubsection{Feature extraction}
            \label{sec:feature_extraction}
            In feature extraction, all parameters in the pre-trained models are frozen when training on the downstream tasks. We utilize the representations extracted from the pre-trained model as fixed features and feed them into a simple, trainable layer. 
            Following the typical setting, we use the representations of the last layer as the features. We also investigate a weighted sum approach proposed by ELMo~\cite{peters2018deep} to fuse representations from various layers rather than the last one in the pre-trained networks and learn the weights along with the prediction layer from downstream tasks.

        \subsubsection{Fine-tuning}
            As for fine-tuning, we also build the classifier with a pre-trained network followed by a simple prediction layer. However, in fine-tuning, the parameters of the entire model are further trained on the downstream tasks. This technique boosts the classifier's performance, especially on difficult downstream tasks such as phoneme classification, but requires longer training time, more task-specific parameters, and a larger memory footprint. 
    \subsection{Probing}
        We also propose probing tasks to understand how knowledge is encoded in pre-trained networks. For the probing tasks, we built classifiers with three different prediction layers: linear, one fully-connected, and two fully-connected layers. We explore several variations of the prediction layers to mitigate the possible bias introduced by network architectures. The prediction layers take representations from each layer of a pre-trained network as the input features and are trained on downstream tasks with the pre-trained model frozen. With the probing, we measure the information richness for each layer's representation based on the classifiers' performance. Such analysis allows us to interpret how information is encoded in the pre-trained networks and provides a new avenue to achieve better performance by using the networks more efficiently.

\begin{table}[t]
  \caption{Hyperparameter for different downstream tasks, LR: Learning rate, SPK: speaker, PH: phoneme}

  \label{tab:hyperparameter}
    \begin{center}
    \begin{small}
  \resizebox{\linewidth}{!}{
    \begin{tabular}{c|ccc}
      \hline
      \textbf{Downstream}  &\textbf{\#  PH / SPK}&\textbf{Detail}&\textbf{LR}\\
      \hline
      \multirow{2}{*}{\begin{tabular}{c}Phoneme\\classification\end{tabular}}    & 72 & weighted-sum     & 1e-3                                     \\ \cline{2-4}
      & 72 & fine-tune                     & 1e-4                   \\
      \hline
      \multirow{2}{*}{\begin{tabular}{c}Utterance-level\\speaker classification\end{tabular}} & 921 & weighted-sum               & 1e-3                                     \\ \cline{2-4}
        & 251   & weighted-sum                & 1e-3   \\
      \hline
    \end{tabular}
  }
  \end{small}
  \end{center}
\end{table}

\section{Experiment results and discussions}
    \subsection{Experimental setup}
        We evaluate the pre-trained networks with one phoneme classification task and three speaker classification tasks as the downstream tasks. In our experiment, we use a 160-dimension acoustic feature, i.e., an 80-dimension log mel-spectrogram and its delta and apply Cepstral Mean and Variance normalization (cmvn), as the input for the pre-trained networks. At the pre-training stage, we train our models with learning rate 5e-5, batch size 50, and 
       AdamW optimizer~\cite{loshchilov2018decoupled} for 500k steps. The models are pre-trained on a single NVIDIA Tesla V100 32GB. We apply different hyperparameters to train classifiers for each downstream task and show the detailed settings in Table~\ref{tab:hyperparameter}. 
        
        We utilize LibriSpeech~\cite{panayotov2015librispeech} for our experiments. LibriSpeech contains three subsets with 500, 360, and 100 hours of speech (denoted as train-other-500, train-clean-360, and train-clean-100) and the transcription and speaker labels. We pre-train our networks on the train-clean-360 set without using any annotation. The train-clean-360 and train-clean-100 sets are used as downstream tasks for evaluating performance in phoneme and speaker classification. The two sets are further split into training, development, and test subset in the ration of 8:1:1 for our experiment. To obtain frame-level phoneme labels to benchmark phoneme classification results, we adopt Montreal Forced Aligner~\cite{mcauliffe2017montreal} to force align transcription to phoneme sequences containing 72 phoneme classes.

\begin{figure}[h]
    \centering
    \begin{tikzpicture}
        \begin{axis}[
            xlabel={Number of layers},
            ybar,
            ymin=97,
            ymax=101,
            ytick={97, 98, 99, 100},
            y label style={
                    at={(axis description cs: 0.5,1.58)},
                    anchor=center,
                    rotate=-90,
                    font=\footnotesize,
                    },
            ylabel={Number of million parameters of AALBERT (A) and Mockingjay (M) },
            extra y ticks={99},
            extra y tick labels = {Accuracy (\%)},
            extra y tick style = {
                            tick label style={
                                            rotate=90,
                                            yshift=5.5mm,},
                            },
            bar width=0.45cm,
            enlarge x limits=0.25,
            height=3.5cm,
            legend cell align={left},
            legend style={
                    at={(0.5,-0.75)},
                    draw=none,
                    /tikz/every even column/.append style={column sep=10pt},
                    anchor=north,
                    legend columns=2
                },
            transpose legend,
            symbolic x coords={3, 6, 12},
            extra x ticks={3, 6, 12},
            extra x tick labels={{7.4 \textsubscript{(A)} 22.6 \textsubscript{(M)}}, {7.4 \textsubscript{(A)}, 44.4 \textsubscript{(M)}}, {7.4 \textsubscript{(A)}, 84.3 \textsubscript{(M)}}},
            extra x tick style={
                    ticklabel pos=top,
                },
            xtick=data,
            x=2.5cm,
            nodes near coords,
            nodes near coords align={vertical},
                    nodes near coords style={
                            font=\footnotesize,
                            /pgf/number format/.cd,
                            fixed,
                            fixed zerofill,
                            precision=1,
                        },
            ]
            \addplot [fill=cyan!50!white]
            	coordinates {(3,99.12) (6, 99.33) (12, 99.33)};
            \addplot [fill=orange!80!white]
            	coordinates {(3, 99.37) (6, 99.54) (12, 99.79)};
            \addplot [fill=teal!60!white,]
            	coordinates {(3, 98.78) (6, 98.94) (12, 97.97)};
            \addplot [fill=red!75!white,]
            	coordinates {(3, 99.37) (6, 99.54) (12, 99.79)};
        \legend{AALBERT (251), Mockingjay (251), AALBERT (921), Mockingjay (921)}
        \end{axis}
    \end{tikzpicture}
    \caption{Speaker classification accuracy on different models and settings with their model parameters. "AALBERT / Mockingjay (251)": settings of utterance-level speaker classification on 251 speaker, "AALBERT / Mockingjay (921)": settings of utterance-level speaker classification on 921 speaker}
    \label{fig:speaker_classification}
\end{figure}
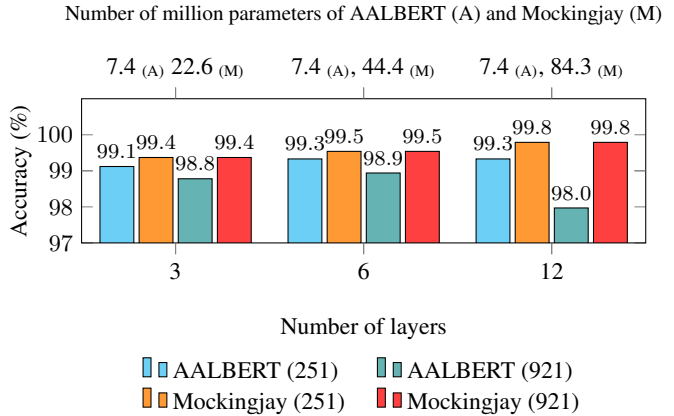

    \subsection{Phoneme classification}
        To measure pre-trained networks' performance on phoneme classification, we build phoneme classifiers with the pre-trained networks followed by two fully-connected layers for prediction. Two pre-trained networks, Mockingjay~\cite{liu2020mockingjay} and AALBERT, are investigated here, with the former as the baseline.

        In Fig~\ref{fig:phoneme_classification}, we show the performance of our models with different layers and settings and compare them to the baseline model (Mockingjay). 
        The vertical axis is the phoneme classification accuracy, while the horizontal axis is the number of network parameters.
        For both fine-tuning and weighted-sum case, AALBERT shows comparable classification accuracy compared to Mockingjay, but with much fewer network parameters. 
        We also note that both 12-layer AALBERT and Mockingjay (denoted as AALBERT-12L and Mockingjay-12L) does not provide performance gain as compared to the 3- and 6- layer counterpart.  We conjecture that the saturation in performance is because we use a limited amount of pre-training data. The 6-layer AALBERT and Mockingjay are sufficient to encode knowledge in our 360 hours of pre-training data.

        In Fig~\ref{fig:weightedsum} and Fig~\ref{fig:finetune}, we show the performance on phoneme classification tasks of both feature-extraction case and fine-tuning case versus different proportions of training data being used. Here are two observations. First of all, not only Mockingjay but AALBERT outperforms the input acoustic feature (shown in Fig~\ref{fig:weightedsum}, Fig~\ref{fig:finetune}). 
        Secondly, these figures show that the representations extracted from Mockingjay and AALBERT have similar performance on phoneme classification tasks.

\subsection{Speaker classification}
Then, we evaluate the model performance with utterance-level speaker classification in train-clean-100 and train-clean-360 subsets. There are 251 and 921 speakers in the two subsets, respectively.
We only use the weighted-sum representations in this part due to space limitation. 

\begin{figure}[h]
    \centering
    \begin{subfigure}[t]{.5\columnwidth}
        \includegraphics[width=\textwidth]{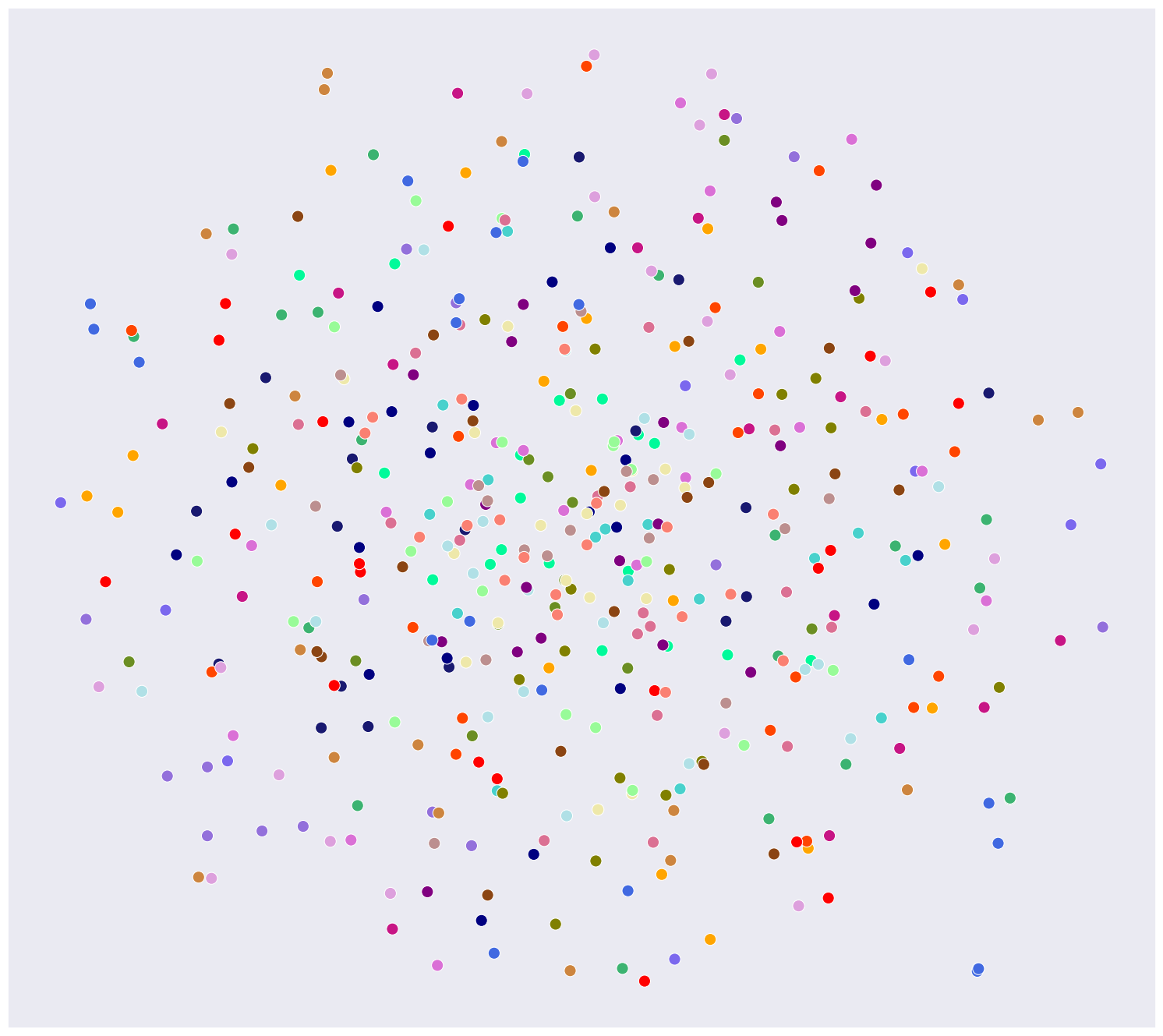}
        \caption{mel and its delta}
        \label{fig:tsne-tradition}
    \end{subfigure}%
    \begin{subfigure}[t]{.5\columnwidth}
        \includegraphics[width=\textwidth]{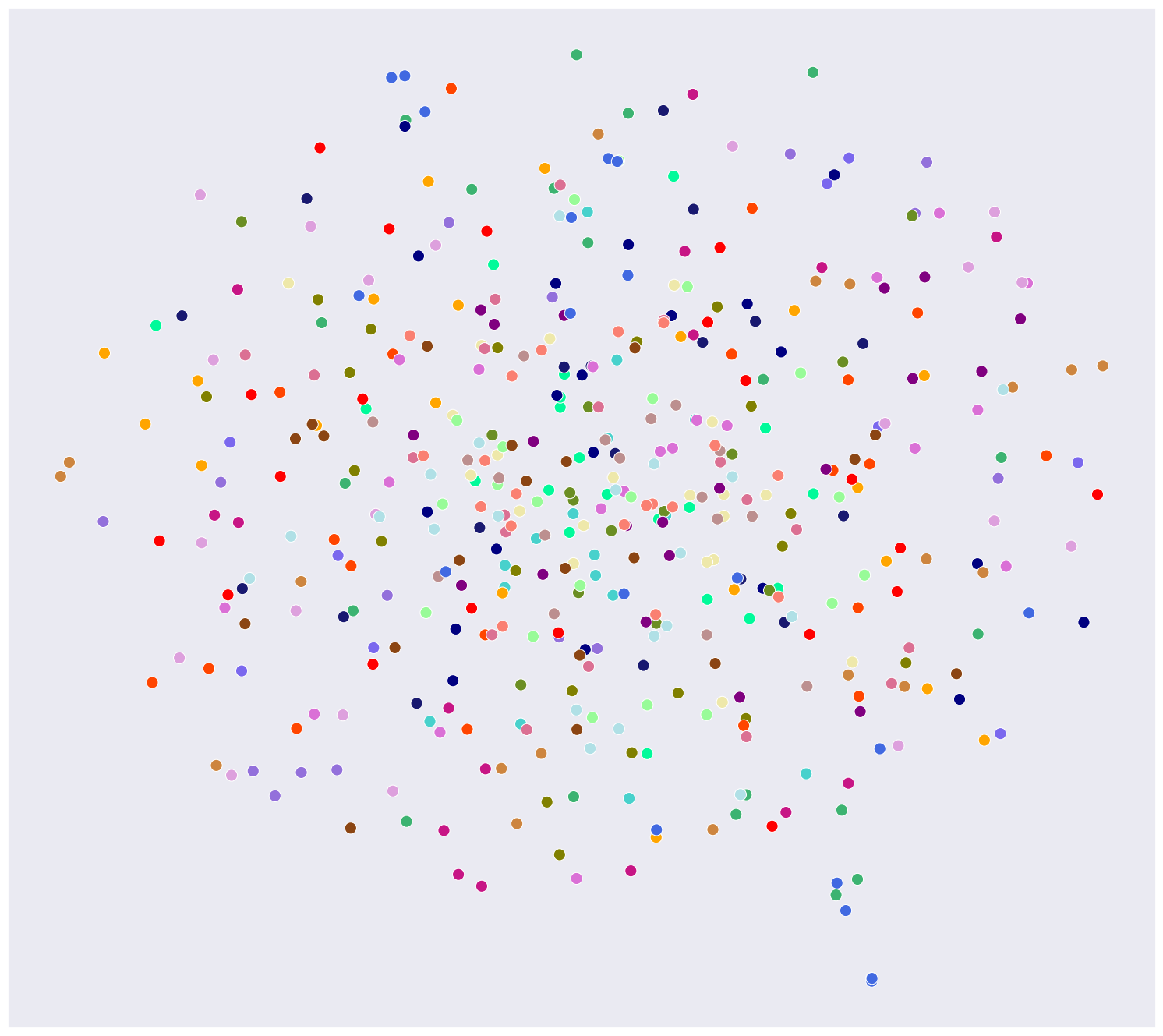}
        \caption{mel spectrogram}
        \label{fig:tsne-tradition2}
    \end{subfigure}
    \centering
   \begin{subfigure}[t]{.75\columnwidth}
        \captionsetup{justification=centering}
       \includegraphics[width=\textwidth]{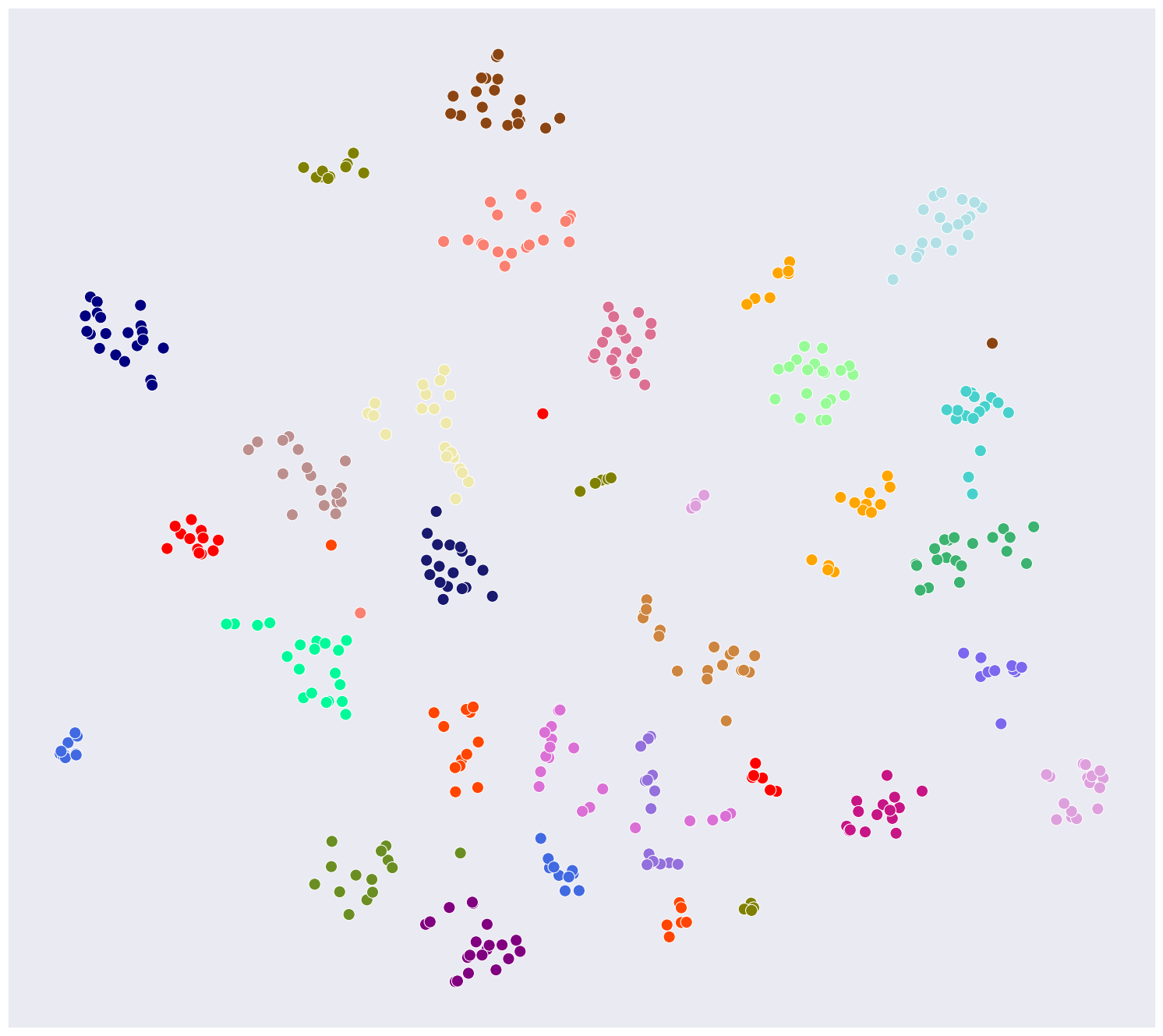}
       \caption{AALBERT representation}
       \label{fig:tsne-aalbert}
   \end{subfigure}
   \caption{Visualization of 25 speakers representations (last layer) via t-SNE in train-clean-100 dataset. Different colors represent different speakers.}

\end{figure}

  On top of the representation, we utilize a linear layer followed by a mean-pooling layer for the classification.
  In Fig~\ref{fig:speaker_classification}, we visualize the performance of speaker classification using Mockingjay and AALBERT representations. The performance of the baseline using the input acoustic feature is not shown in the figure, where the accuracy is 0.6\%. The results suggest that both AALBERT and Mockingjay encode more abundant speaker information than the raw input of the two pre-trained networks and yield nearly perfect classification results, while AALBERT leverages much fewer parameters than Mockingjay.

  Furthermore, we use t-SNE~\cite{maaten2008visualizing} to visualize the utterance representations extracted from input acoustic feature and AALBERT in Fig~\ref{fig:tsne-tradition} and Fig~\ref{fig:tsne-aalbert}. 
  In the figures, each point represents an utterance with its embeddings generated by mean-pooling; we encoder each speaker with a different color.
  The utterance representations from AALBERT for each speaker are clustered together, while we cannot observe the same phenomenon from the input acoustic features. The result shows that AALBERT better encodes speaker information.

In conclusion, AALBERT shows comparable results on speaker classification tasks against Mockingjay, yet using  $91\%$ fewer parameters.

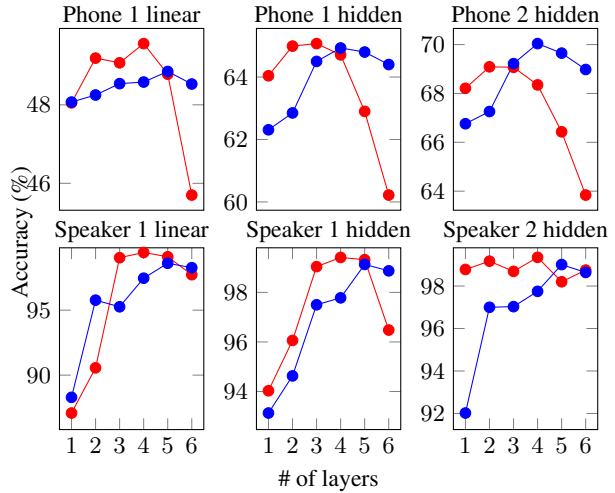
\begin{figure}[th]
    \pgfplotsset{
        every axis title/.append style={
                at={(0.5, 0.9)},
                font=\small,
            }
    }
    \begin{tikzpicture}
        \begin{groupplot}[
                group style={
                        group size=3 by 2,
                        vertical sep=0.5cm, horizontal sep=0.7cm,
                    },
            ]
            \nextgroupplot[
                title={Phone 1 linear},
                ylabel=Accuracy (\%),
                ylabel style={
                        yshift=-3mm,
                        xshift=-1.5cm,
                    },
        		width=3.5cm,
        		height=4cm,
        		xtick=\empty,
        	]
        	\addplot[color=red,mark=*] coordinates {
        		(1,48.05)
        		(2,49.19)
        		(3,49.07)
        		(4,49.56)
        		(5,48.78)
        		(6,45.70)
        	};
        	
        	\addplot[color=blue,mark=*] coordinates {
        		(1,48.07)
        		(2,48.25)
        		(3,48.54)
        		(4,48.58)
        		(5,48.85)
        		(6,48.53)
        	};
            \nextgroupplot[
        		width=3.5cm,
        		height=4cm,
        		xtick=\empty,
                title={Phone 1 hidden},
                ]
        	\addplot[color=red,mark=*] coordinates {
        		(1,64.04)
        		(2,64.99)
        		(3,65.07)
        		(4,64.71)
        		(5,62.90)
        		(6,60.22)
        	};
        	
        	\addplot[color=blue,mark=*] coordinates {
        		(1,62.31)
        		(2,62.85)
        		(3,64.50)
        		(4,64.93)
        		(5,64.80)
        		(6,64.40)
        	};
        	
            \nextgroupplot[
                title={Phone 2 hidden},
        		width=3.5cm,
        		height=4cm,
        		xtick=\empty,
        	]
        	\addplot[color=red,mark=*] coordinates {
        		(1,68.21)
        		(2,69.09)
        		(3,69.07)
        		(4,68.35)
        		(5,66.43)
        		(6,63.84)
        	};
        	
        	\addplot[color=blue,mark=*] coordinates {
        		(1,66.76)
        		(2,67.26)
        		(3,69.22)
        		(4,70.04)
        		(5,69.65)
        		(6,68.98)
        	};
        \nextgroupplot[
        		width=3.5cm,
        		height=4cm,
        		ymax=99.8,
        		xtick=data,
                title={Speaker 1 linear},
        	    ]
        	\addplot[color=red,mark=*] coordinates {
        		(1,87.08)
        		(2,90.56)
        		(3,99.04)
        		(4,99.44)
        		(5,99.11)
        		(6,97.73)
        	};
        	
        	\addplot[color=blue,mark=*] coordinates {
        		(1,88.29)
        		(2,95.77)
        		(3,95.26)
        		(4,97.46)
        		(5,98.61)
        		(6,98.27)
        	};

            \nextgroupplot[
                xlabel=\# of layers,
        		ymax=99.8,
        		width=3.5cm,
        		height=4cm,
        		xtick=data,
                title={Speaker 1 hidden},
        	]
        	\addplot[color=red,mark=*] coordinates {
        		(1,94.03)
        		(2,96.06)
        		(3,99.04)
        		(4,99.41)
        		(5,99.32)
        		(6,96.48)
        	};
        	
        	\addplot[color=blue,mark=*] coordinates {
        		(1,93.13)
        		(2,94.63)
        		(3,97.50)
        		(4,97.78)
        		(5,99.13)
        		(6,98.87)
        	};
            \nextgroupplot[
        		width=3.5cm,
        		height=4cm,
        		xtick=data,
        		ymax=99.8,
                title={Speaker 2 hidden},
        	]
        	\addplot[color=red,mark=*] coordinates {
        		(1,98.78)
        		(2,99.17)
        		(3,98.69)
        		(4,99.35)
        		(5,98.20)
        		(6,98.75)
        	};
        	
        	\addplot[color=blue,mark=*] coordinates {
        		(1,92.02)
        		(2,97.00)
        		(3,97.03)
        		(4,97.75)
        		(5,99.01)
        		(6,98.64)
        	};
        \end{groupplot}
    \end{tikzpicture}
\caption{Probing task, Blue Line: AALBERT-6L, Red line: Mockingjay-6L}
\label{fig:Probing}
\end{figure}

\begin{figure}[t]
    \centering
    \begin{subfigure}{.45\columnwidth}
        \includegraphics[width=\textwidth]{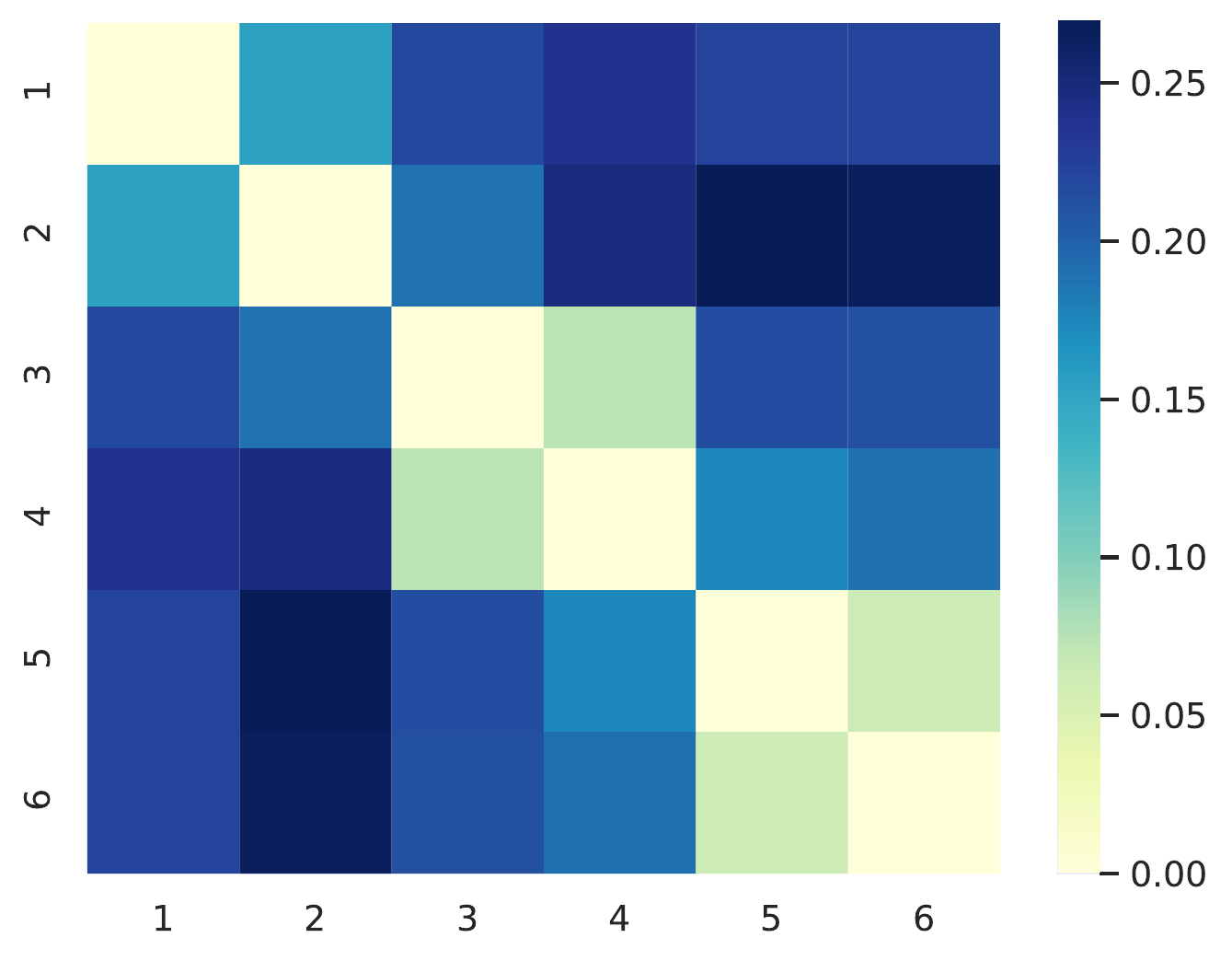}
        \caption{
        Average}
        \label{fig:js_attention_aalbert}
    \end{subfigure}%
    \begin{subfigure}{.45\columnwidth}
        \begin{minipage}[t]{.5\columnwidth}
            \includegraphics[width=\textwidth]{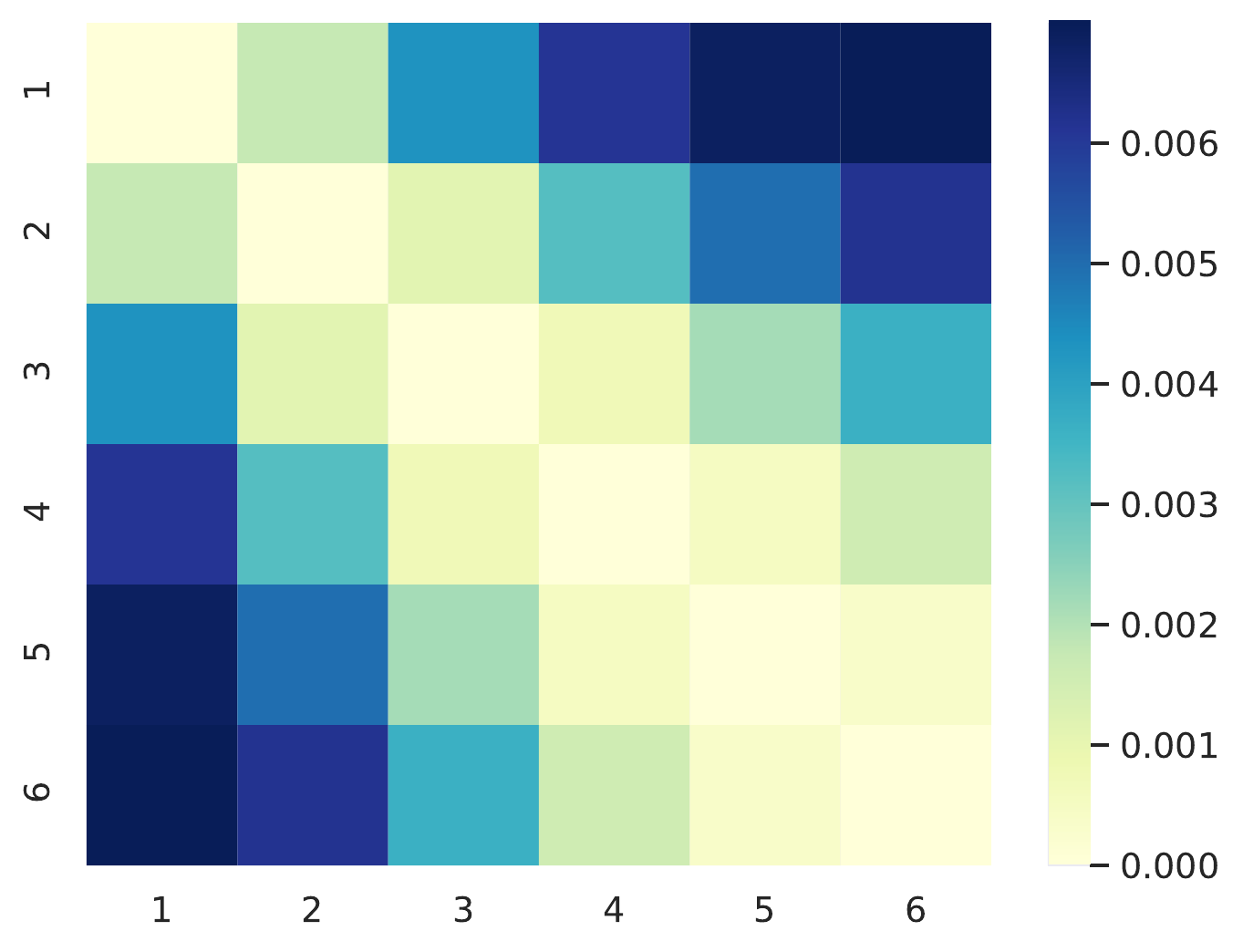}
            \caption{2\textsuperscript{nd} head}
            \label{fig:js_attention_aalbert_1}
        \end{minipage}%
        \begin{minipage}[t]{.5\columnwidth}
            \includegraphics[width=\textwidth]{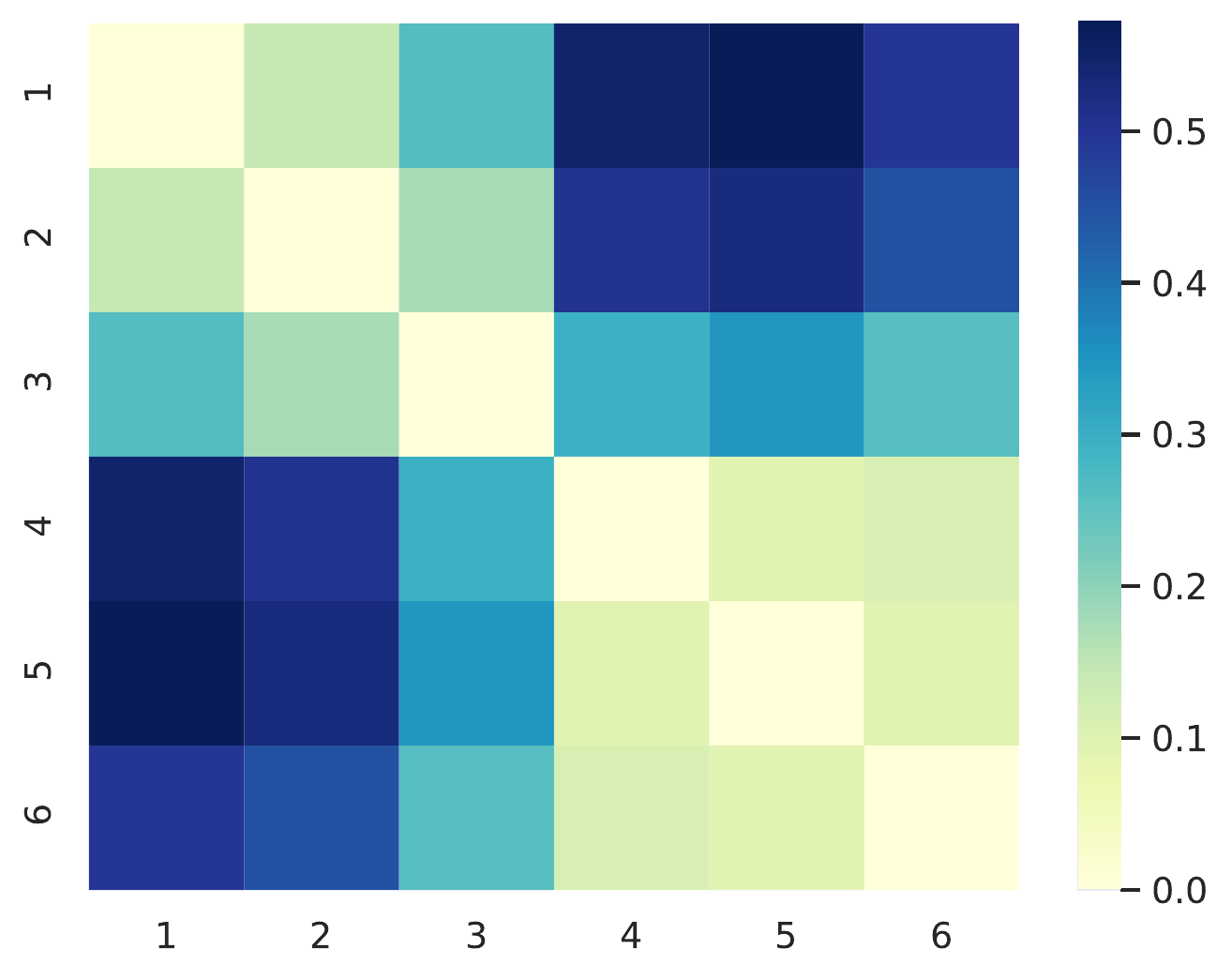}
            \caption{8\textsuperscript{th} head}
            \label{fig:js_attention_aalbert_7}
        \end{minipage}
        \begin{minipage}[t]{.5\columnwidth}
            \includegraphics[width=\textwidth]{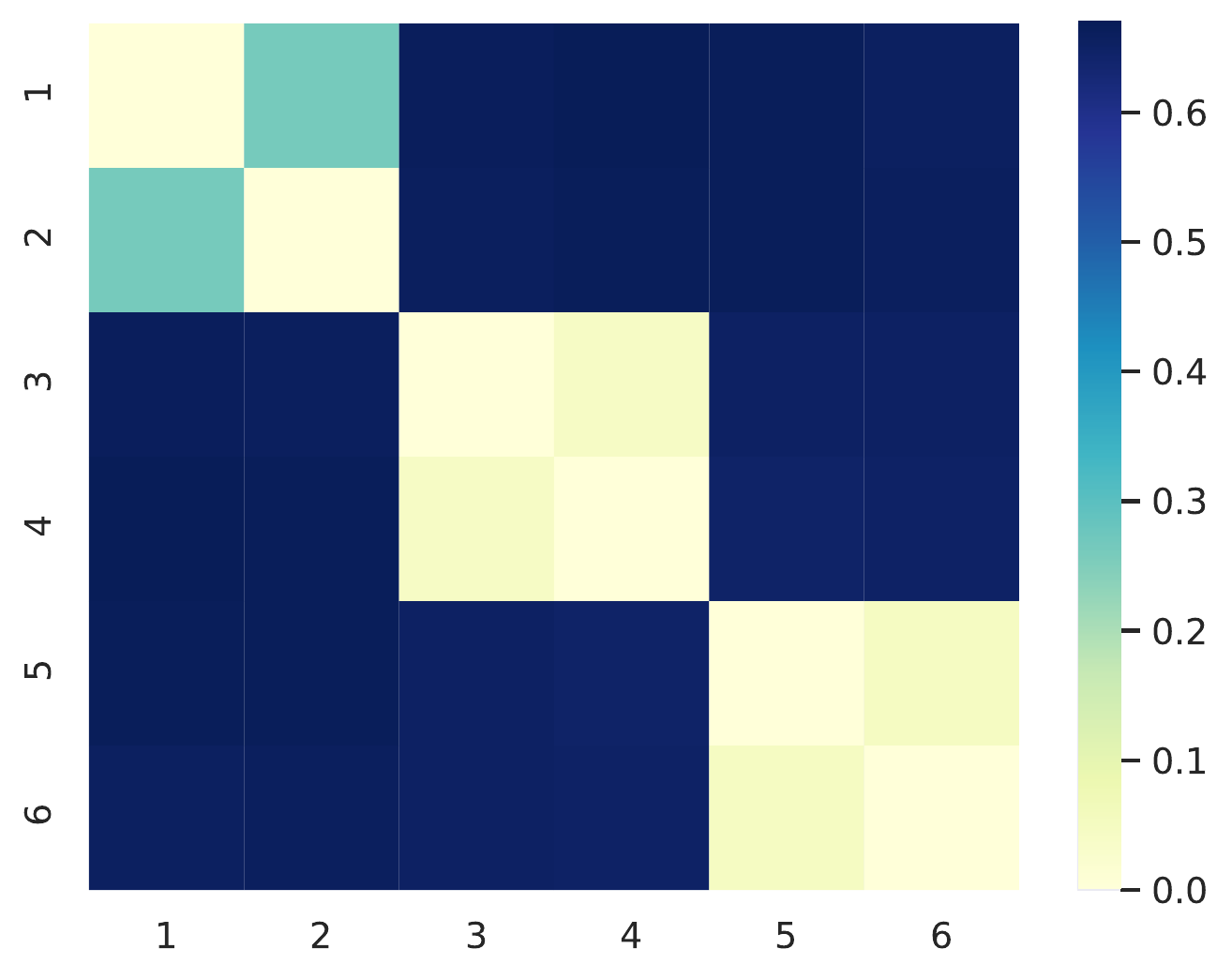}
            \caption{11\textsuperscript{th} head}
            \label{fig:js_attention_aalbert_10}
        \end{minipage}%
        \begin{minipage}[t]{.5\columnwidth}
            \includegraphics[width=\textwidth]{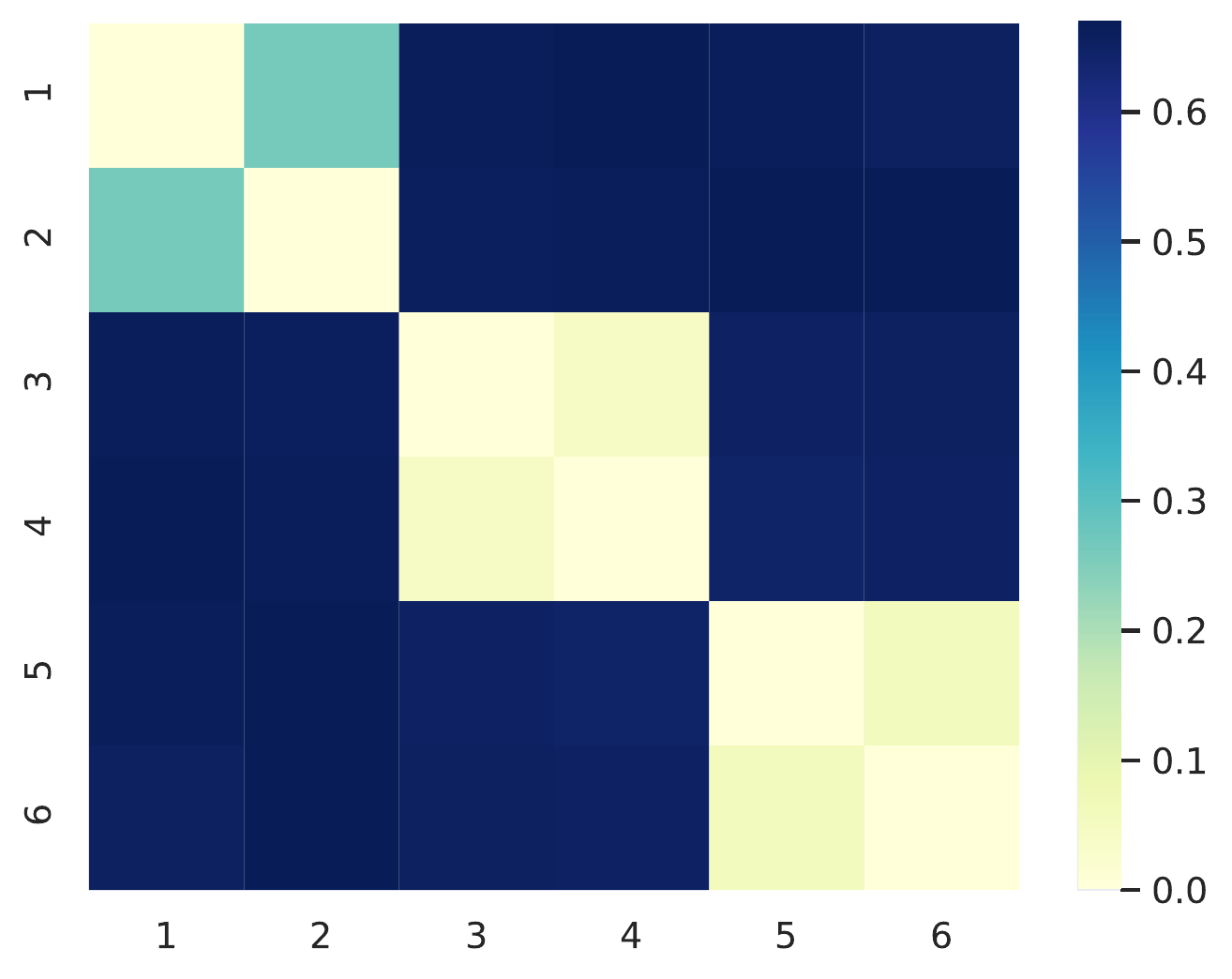}
            \caption{12\textsuperscript{th} head}
            \label{fig:js_attention_aalbert_11}
        \end{minipage}
    \end{subfigure}
    \caption{The JS divergence of attention distribution 
    between different layers in AALBERT-6L. (a) is the average case, while (b)(c)(d)(e) represent different attention heads.}
    \label{fig:different_js_aalbert}
\end{figure}

\subsection{Probing task}
  We utilize two probing tasks, phoneme classification and frame-level speaker classification\footnote{Since we want to analyze an individual representation instead of the whole utterance, we choose frame-level instead of utterance-level.}, to examine how much phoneme and speaker information contain in the representations of each layer. 
  In both tasks, we use train-clean-100 dataset, which is unseen at the pre-training stage.
  We probe AALBERT-6L and Mockingjay-6L since the average performances of them are the best. 
 We utilize three different classifiers as the probing models, linear, one hidden layer, and two hidden layers, to probe each layer of the pre-trained models for the speaker information and the phoneme information.
  We use several probing models with different network architectures to mitigate the possible bias from the probing models.

 Fig~\ref{fig:Probing} shows the result of probing tasks. 
 For the probing of phoneme information, the three different probing models show the same trends among the same pre-training model.
 In both pre-training models, as the depth increases, the phoneme information increases first and then decreases.
 Comparing the two pre-training models, the peak of the Mockingjay-6L is closer to the input layers than AALBERT-6L.
 On the other hand, when comparing the absolute performance of Mockingjay-6L and AALBERT-6L, the conclusion from different probing models would be different.
 Mockingjay-6L achieves better phoneme classification accuracy for the shallower probing model, whereas AALBERT-6L obtains better performance of the deeper probing model.
 For speaker information, the $5$\textsuperscript{th} layer of AALBERT-6L contains the most speaker information, while the $4$\textsuperscript{th} layer is the best for Mockingjay-6L.
 
 The results in Fig~\ref{fig:Probing} further indicate that the intermediate representations outperform the representations from the last layer in all four different probing tasks regardless of Mockingjay-6L or AALBERT-6L model.
 This might indicate that the last layer fits the pre-training tasks too much; therefore, the representations extracted from the intermediate layers may be more suitable for downstream tasks.

\subsection{Attention distribution in AALBERT}
        Here we repeat section~\ref{sec:mockingjay} experiments on AALBERT-6L.
        Fig~\ref{fig:js_attention_aalbert} shows that the JS divergence of attention distribution is very small between layer 1, 2, layer 3, 4, and layer 5, 6.
        Fig~\ref{fig:js_attention_aalbert_1} shows that the JS divergence are small in diagonal and its neighbor area.
        On the contrary, the first and the last layer differ a lot.
        In Fig~\ref{fig:js_attention_aalbert_7}, the JS divergence between first three layers are small, and so do the last three ones.
        However, the JS divergence between these two parts are large. In Fig~\ref{fig:js_attention_aalbert_10},~\ref{fig:js_attention_aalbert_11}, the JS divergence of layer 1,2, layer 3,4 and layer 5,6 
        are small, but the JS divergences between every two layers of them are large. 
        These results show that the same parameters may still cause totally different attention distribution over different layers.

\section{Acknowledgement}
We thank to National Center for High-performance Computing (NCHC) for providing computational and storage resources.

\section{Conclusion}
 In this paper, we present a novel model, Audio ALBERT (AALBERT). AALBERT is a pre-trained model for extracting latent representations that encode the audio information. The model is learned by reconstructing the masked input acoustic features to the linear spectrogram. We show that AALBERT can achieve comparable performances against Mockingjay, a BERT-like pre-trained audio model, yet with much fewer parameters. Besides, we show promising results in encoding audio information with much smaller pre-trained models. 
 For our future work, we will investigate various model architectures to improve further the efficiency of pre-trained models in computation and parameter usage.

 \bibliographystyle{ieee}
 \bibliography{main}

\end{document}